\documentclass[a4paper,fleqn]{cas-sc}

\usepackage[numbers]{natbib}
\usepackage{graphicx}
\usepackage{subcaption}

\def\tsc#1{\csdef{#1}{\textsc{\lowercase{#1}}\xspace}}
\tsc{WGM}
\tsc{QE}

\begin{document}
\let\WriteBookmarks\relax
\def\floatpagepagefraction{1}
\def\textpagefraction{.001}

\shorttitle{}    

\shortauthors{}  

\title [mode = title]{Non-classical photon statistics in frequency-modulated double quantum dot cavity systems}  

\author[1]{Tatiana Mihaescu}

\affiliation[1]{organization={National Institute of Physics and Nuclear Engineering},
            addressline={Reactorului 30}, 
            city={M\u{a}gurele-Bucharest},
         postcode={077125}, 
            country={Romania}}
            
\author[1]{Aurelian Isar}

\author[2]{Mihai A. Macovei}
\cormark[1]
\ead{mihai.macovei@ifa.usm.md}

\affiliation[2]{organization={Institute of Applied Physics, Moldova State University},
            addressline={Academiei str. 5}, 
            city={Chi\c{s}in\u{a}u},
            postcode={MD-2028}, 
            country={Moldova}}

\cortext[1]{Corresponding author}

\begin{abstract}
The quantum dynamics of a double quantum dot two-level system, coupled to a leaking microwave resonator mode, is 
theoretically investigated. The double quantum dot is driven by applying a continuous field giving rise to time-dependent 
modulations of its energy separation between the ground levels of each dot, forming the qubit. Therefore, 
the cavity resonances occur when the difference between the resonator and the qubit frequencies equals the multiple of the 
modulation frequency, respectively. In the strong Coulomb interaction limit and the weak couplings of the combined 
system, i.e. the qubit and the single-mode cavity, to corresponding electronic, phononic or photonic reservoirs, we have 
demonstrated that the output cavity electromagnetic field is formed of a single-photon flux obeying the sub-Poissonian 
quantum photon statistics. This was proved via Fano's factor behaviour or by comparing the resonator's second- and 
third-order photon correlation functions, respectively. The phonon, the cavity mode dephasing or the environmental 
temperatures influence on the quantum properties of the photons are also discussed.
\nocite{*}
\end{abstract}




\maketitle
\section{Introduction}
The emergence of non-classical light from the quantum electron transport in a two level system serves as an essential  toolbox 
for the exploration of fundamental quantum physics and the development of quantum technological applications \cite{rad, zh, su}. 
Conversely, when the cavity is in resonance with the energy splitting of the two-level system, the itinerant microwave photons are 
converted into the charge current flow, and the transport devices become efficient detectors of quantum light \cite{soq, abd, opp}. 
Furthermore, the microwave resonator represents the indispensable resource for long-range coherent interaction between multiple 
qubits by probing the characteristics of the two-level systems and mediating the creation of quantum correlations between them 
\cite{and}. While the coupling with a cavity provides a dynamical control over the qubit by enabling the mutual exchange and 
correlations, a periodic external driving allows for an active, direct and deterministic control over the investigated system 
\cite{td1,grif,ashh,silv}.

Originally, the atomic systems embedded in an optical cavity were the representative setup for the light-matter interaction, 
resulting in the generation of new hybridized states of light as a distinctive feature. Nowadays,  a great  interest is directed 
toward harnessing quantum mechanical effects in the solid-state cavity quantum electrodynamics (QED) platforms featuring 
systems like quantum dots, their advantages being the scalability and the support of strong nonlinearities at single-photon 
level \cite{buc, zhag,lir}.  Especially suitable candidates for the charge qubit-like systems are the double quantum dots (DQDs), 
since they support a coherent electron dynamics between dots and induce a strong dipole moment, which allows for a strong 
coupling regime with the photons in the microwave cavities \cite{rev_art,fre,liu,khan,hal,abdul}. This configuration perfectly suits 
the highly desirable implementation of applications which exploit nonclassical states of light by means of an electronic interface 
for the manipulation of photon statistics \cite{xu,ye,wang,nian}. Likewise, the electrical current through the double quantum 
dot and the microwave field inside the resonator show interdependent mutual critical steady-state behaviours, suggesting 
ways to design on-chip microwave quantum switching devices for instance \cite{tam1,tam2}, see also Ref.~\cite{dqph}.

The implementation of an additional external coherent drive bears the potential to transform the dipole-coupled 
DQD-cavity system into a highly tunable nonlinear quantum platform \cite{td4,td5}. For instance, it enables the Floquet 
engineering of the DQD, such that the energy spectrum can be tailored to match the effective resonance conditions with 
the cavity \cite{shirai, bert}.  This provides an independent control of transition pathways in the DQD, which may induce 
the Landau–Zener–Stückelberg interference phenomena where the transition sweeps through avoided crossings due to 
the periodic drive \cite{td2}.  Though the electron-phonon coupling appearing naturally in the semiconducting DQDs and 
acting as a relaxation mechanism of the qubit, the periodic drive may still produce an effective population inversion of 
the Floquet states. This resulted in a cavity power gain and enhanced photon emission \cite{bert, gull}. Consequently, 
the emergence of non-classical states of light in diverse regimes can be adjusted based on the driving parameters, 
and the microwave photons were observed to be emitted even in the absence of a dc current \cite{steh}. Thus, in the 
context of open quantum systems, where the periodically driven systems are also subjected to dissipations which are 
ubiquitous in nature, the strong external driving enables the dynamical control of steady-state properties beyond 
equilibrium \cite{kohler,cao}.

Here, we employ a microscopic theoretical model in order to investigate a hybrid compound sample consisting of a 
double quantum dot (DQD) two-level emitter interacting with a leaking cavity mode. Therewith the DQD's frequency 
is being time-modulated by applying a microwave continuous electromagnetic field. We have obtained the master 
equation describing this complex system, where the pumping due to electronic reservoirs as well as phonon or 
surrounding temperatures influences on DQD plus cavity photon subsystems quantum dynamics are taken into 
account, respectively. The derived master equation allowed us to obtain the expectation values of the cavity photon 
numbers as well their second- or third-order correlation functions, including Fano's factor. Hence, at lower 
environmental temperatures, we have demonstrated that the cavity output electromagnetic field, in resonance with 
modulated frequencies around the qubit one, consists of a stream of single photons possessing sub-Poissonian 
photon statistics. The presence of phonons in the sample may diminish the effect depending, respectively, on the 
interplay between the phonon and electronic pumping rates. However, via the engineering of the environmental 
phonon mode reservoir, their influence may be reduced or even cancelled.

This paper is organized as follows. In Sec.~\ref{theo} we describe the system of interest and the analytical approach, while 
in Sec.~\ref{seqm} we represent the equations of motion characterising the discussed system, respectively.  Sec.~\ref{RD} 
presents and analyses the obtained results. The article concludes with a summary given in Sec.~\ref{sum}.

\section{The model and theoretical approach \label{theo}}
\begin{figure}
\centering
\includegraphics[width =10cm]{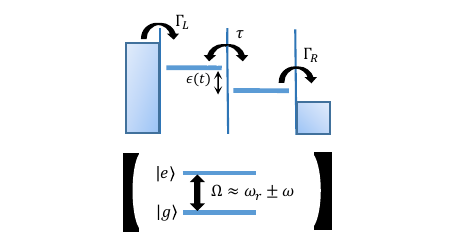}
\caption{\label{fig-0}
Schematic picture of the system. The resonator frequency $\omega_{r}$ is being near resonance with the 
modulated DQD frequency $\Omega$. Electron's tunnelling processes are illustrated by solid arrows.} 
\end{figure}

We consider a double quantum dot (DQD) coupled to the single mode of a leaking microwave resonator, see Fig.~(\ref{fig-0}). 
The DQD is connected by tunnel junctions to two large reservoirs $L$ (left) and $R$ (right) and, within the strong Coulomb 
interaction limit, electrons can only enter the DQD system from the left and leave it only to the right, respectively. Therefore, 
the DQD is restricted to three possible configurations \cite{rev_art}: the null-electron subspace or the empty-dot state, i.e. 
$|0\rangle$, together with the single-electron subspaces, where the electron is localized either on the left, $|L\rangle$, or the 
right dot, $|R\rangle$, with $|L\rangle\langle L|+|R\rangle\langle R| + |0\rangle\langle 0|=1$. A time-dependent oscillating 
signal is applied then so that the time-dependent frequency difference becomes, see e.g. \cite{td1,td4,td2,td3}: 
$\epsilon(t) = \epsilon + \epsilon_{0}\cos{(\omega t + \phi)}$, where the amplitude $\epsilon_{0}$ of the applied field of 
frequency $\omega$ and phase $\phi$ is being considered smaller than the frequency separation 
$\epsilon$ in the absence of the external driving, i.e., $\epsilon_{0} \ll \epsilon$, see Fig.~(\ref{fig-0}). 

Hence, the Hamiltonian describing this complex system is given as follows: $H=H_{1} + H_{2}$, where 
$H_{1}=H_{q} + H_{eL} + H_{eR} + H_{pn}$, while $H_{2}=H_{r} + H_{pt}$. Correspondingly,
\begin{eqnarray}
H_{q} = \frac{\hbar\epsilon(t)}{2}\bigl(|L\rangle\langle L| - |R\rangle\langle R| \bigr) +\hbar\tau\bigl(|L\rangle\langle R| 
+ |R\rangle\langle L|\bigr), \label{hq} 
\end{eqnarray}
is the Hamiltonian of two quantum dots, forming the DQD, with $\hbar\epsilon(t)$ being their energy separation, while 
$\tau$ is the inter-dot tunnelling amplitude, see Fig.~(\ref{fig-0}). The DQD interaction with the two 
fermionic leads, $L$ and $R$ maintained at different chemical potentials, is characterized by the following Hamiltonians
\begin{eqnarray}
H_{eL} &=& \hbar\sum_{\chi_{L}}\bigl\{\omega_{\chi_{L}}c^{\dagger}_{\chi_{L}}c_{\chi_{L}} + 
g_{\chi_{L}}\bigl(c_{\chi_{L}}|L\rangle\langle 0|  + |0\rangle\langle L|c^{\dagger}_{\chi_{L}}\bigr)\bigr\}, 
\label{heL}
\end{eqnarray}
and
\begin{eqnarray}
H_{eR} &=& \hbar\sum_{\chi_{R}}\bigl\{\omega_{\chi_{R}}c^{\dagger}_{\chi_{R}}c_{\chi_{R}} 
+ g_{\chi_{R}}\bigl(c_{\chi_{R}}|R\rangle\langle 0|  + |0\rangle\langle R|c^{\dagger}_{\chi_{R}}\bigr)\bigr\}. 
\label{heR}
\end{eqnarray}
In the DQD-leads Hamiltonians, $c^{\dagger}_{\chi_{\alpha}}(c_{\chi_{\alpha}})$ is the creation (annihilation) operator 
for fermions with wave vector $\chi$ in the $\alpha$th electrode, whereas $g_{\chi_{\alpha}}$ is the coupling constant 
between the DQD and the fermionic baths, $\{\alpha \in L,R\}$. The electron–phonon interaction, governed by the 
material’s properties, is characterized by the next Hamiltonian, i.e.,
\begin{eqnarray}
H_{pn} &=& \hbar\sum_{p}\bigl\{\omega_{p}b^{\dagger}_{p}b_{p} + g_{p}\bigl(|L\rangle\langle L| 
- |R\rangle\langle R|\bigr) \bigl(b_{p}+ b^{\dagger}_{p}\bigr)\bigr\},  \label{hpn}
\end{eqnarray}
where the first component is the free energy of the phonon reservoir, with $b^{\dagger}_{p}$ and $b_{p}$ being the 
generation and annihilation phonon operators satisfying the boson commutation relations, $[b_{p},b^{\dagger}_{p'}]=
\delta_{pp'}$ and $[b_{p},b_{p'}]=[b^{\dagger}_{p},b^{\dagger}_{p'}]=0$, while the second one accounts for the 
DQD–phonon interaction with $g_{p}$ being the corresponding coupling strength. Respectively, the Hamiltonian
\begin{eqnarray}
H_{r} = \hbar\omega_{r}a^{\dagger}a + \hbar g_{r}\bigl(|L\rangle\langle L| - |R\rangle\langle R|\bigr)\bigl(a 
+ a^{\dagger}\bigr), \label{hr}
\end{eqnarray}
describes the free energy of the microwave resonator of frequency $\omega_{r}$ and qubit–cavity interaction, with 
$g_{r}$ denoting the corresponding coupling constant, see Fig.~(\ref{fig-0}). The cavity photon creation (annihilation) operators, 
$a^{\dagger}(a)$, obey the standard commutation relations, i.e., $[a,a^{\dagger}]=1$ and 
$[a,a]=[a^{\dagger},a^{\dagger}]=0$. Furthermore, the resonator is being damped by the action of the environmental 
thermal electromagnetic field (EMF) reservoir while the damping Hamiltonian is given by
\begin{eqnarray}
H_{pt} &=& \hbar\sum_{k}\bigl\{\omega_{k}a^{\dagger}_{k}a_{k} + ig_{k}(a^{\dagger} + a)(a^{\dagger}_{k} - a_{k})\bigr\}. 
\label{hpt}
\end{eqnarray}
Here the thermal bath photon annihilation and creation operators $\{a_{k},a^{\dagger}_{k}\}$ satisfy the same 
bosonic commutation relations as those for phonon operators and, again, the first term corresponds to the 
free energy of the EMF reservoir whereas the second one denotes the interaction of the resonator mode with 
its surrounding thermal bath which is described by the coupling constant $g_{k}$, respectively.

The system is conveniently described if (i) one diagonalizes the unperturbed Hamiltonian (\ref{hq}), using the 
transformation,
\begin{eqnarray}
|L\rangle=\cos{(\theta/2)}|e\rangle-\sin{(\theta/2)}|g\rangle, \nonumber \\
|R\rangle=\sin{(\theta/2)}|e\rangle + \cos{(\theta/2)}|g\rangle, \label{dst}
\end{eqnarray}
with $\cos{\theta}=\epsilon/\Omega$, $\sin{\theta}=2\tau/\Omega$ and $\Omega=\sqrt{\epsilon^{2}+(2\tau)^{2}}$, and (ii)
represent the Hamiltonians (\ref{hq}-\ref{hpt}) via new qubit quasispin operators $R_{\alpha\beta}=|\alpha\rangle \langle\beta|$,
satisfying the commutation relations $[R_{\alpha\beta},R_{\beta'\alpha'}]$=$\delta_{\beta\beta'}R_{\alpha\alpha'}$ - 
$\delta_{\alpha'\alpha}R_{\beta'\beta}$, with $\{\alpha, \beta \in e, g, 0\}$, respectively. Then, in the interaction picture given by 
the unitary operator, 
\begin{eqnarray*}
U(t)=\exp\biggl\{\frac{i}{\hbar}\int^{t}_{0}d\tau H_{0}(\tau)\biggr\},
\end{eqnarray*}
where 
\begin{eqnarray*}
H_{0} &=&\hbar\bigl(\Omega + \epsilon_{0}\cos{\theta}\cos{(\omega t +\phi)}\bigr)R_{z}/2 
+ \sum_{p}\hbar\omega_{p}b^{\dagger}_{p}b_{p} 
+ \sum_{l \in \{L,R\}}\sum_{\chi_{l}}\hbar\omega_{\chi_{l}}c^{\dagger}_{\chi_{l}}c_{\chi_{l}} \nonumber \\
&+& \hbar\omega_{r}a^{\dagger}a + \sum_{k}\hbar\omega_{k}a^{\dagger}_{k}a_{k}, 
\end{eqnarray*}
with $R_{z}=|e\rangle\langle e|-|g\rangle\langle g|$, one transforms the entire system's Hamiltonian in the interaction 
representation. 

Next, assuming that the corresponding resonator as well as DQD interaction with the photonic, electronic or phonon 
environmental reservoirs are weak, one can eliminate these degrees of freedom in the Born–Markov approximations
\cite{gsag,wm}. In this regard, one inserts separately each of the Hamiltonians from (\ref{heL}-\ref{hpn}) and (\ref{hpt}), 
all in the interaction picture, in the time evolution equation for the density matrix operator $\rho$, i.e.,
\begin{eqnarray}
\dot \rho(t) = -\frac{1}{\hbar^{2}}\int^{t}_{0}dt'{\rm Tr}\bigl\{\bigl[H_{I}(t),[H_{I}(t'),\rho(t')]\bigr]\bigr\}, \label{meq}
\end{eqnarray}
where the overdot means differentiation with respect to time, whereas the notation ${\rm Tr\{\cdots\}}$ denotes the trace 
over the corresponding electronic, phonon or photon degrees of freedom. This will allow us to derive the final master 
equation describing the qubit plus leaking resonator mode subsystems only.

Particularly in this regard, the interaction picture Hamiltonian of the DQD-left-lead is represented as follows:
\begin{eqnarray}
H^{(I)}_{eL}&=&\hbar\sum_{\chi_{L}}\sum_{m}g_{\chi_{L}}J_{m}(\eta)\bigl(c_{\chi_{L}}\bigl\{\cos{(\theta/2)}R_{e0}e^{i\Omega t/2} 
e^{im(\omega t + \phi)-i\bar \phi/2} - \sin{(\theta/2)}R_{g0}e^{-i\Omega t/2} \nonumber \\
&\times&e^{-im(\omega t + \phi)+i\bar \phi/2}\bigr\}e^{-i\omega_{\chi_{L}}t} + H.c. \bigr), \label{heLI}
\end{eqnarray}
where we have used the notations $\eta=(\epsilon_{0}/2\omega)\cos{\theta}$, $\bar \phi=2\eta\sin\phi$ as well as the identity 
$\exp{\{\pm i\eta\sin{(\omega t + \phi)}\}}=\sum^{\infty}_{m=-\infty}J_{m}(\eta)\exp{\{\pm im(\omega t +\phi)\}}$ with $J_{m}(\eta)$ 
being the ordinary Bessel function. Inserting (\ref{heLI}) in (\ref{meq}) and performing the trace over the left electronic reservoir, in 
the Born–Markov approximations, one obtains
\begin{eqnarray}
\dot \rho_{eL} &=& -\frac{1}{2}\sum_{mm'}\Gamma_{L}(\Omega'_{m'})J_{m}(\eta)J_{m'}(\eta)\bigl\{\cos^{2}{(\theta/2)}
[R_{0e},R_{e0}\rho]e^{-i(m-m')(\omega t + \phi)} \nonumber \\
&+& \sin^{2}{(\theta/2)}[R_{0g},R_{g0}\rho]e^{i(m-m')(\omega t + \phi)} \bigr \}+ H.c. \label{reL}
\end{eqnarray}
Similarly, the contribution to the entire master equation arising from the DQD-right-lead interaction is
\begin{eqnarray}
\dot \rho_{eR} &=& -\frac{1}{2}\sum_{mm'}\Gamma_{R}(\Omega'_{m'})J_{m}(\eta)J_{m'}(\eta)\bigl\{\sin^{2}{(\theta/2)}
[R_{e0},R_{0e}\rho]e^{i(m-m')(\omega t + \phi)} \nonumber \\
&+& \cos^{2}{(\theta/2)}[R_{g0},R_{0g}\rho]e^{-i(m-m')(\omega t + \phi)} \bigr \} + H.c. \label{reR}
\end{eqnarray}
Here $2\Gamma_{l}(\Omega'_{m})=\pi\sum_{\chi_{l}}g^{2}_{\chi_{l}}\delta(\omega_{\chi_{l}}-\Omega'_{m})$, $\{l \in L,R\}$,
$\Omega'_{m}=\Omega/2 + m\omega$, with $|2m\omega| \ll \Omega$, and we have taken into account that the applied 
bias voltage is much larger than the temperature and the energy difference between the states in resonance.

In the same vein, the DQD-phonon-interaction contribution to the whole master equation is given by
\begin{eqnarray}
\dot \rho_{pn} &=& -\frac{1}{2}\sin^{2}{\theta}\sum_{mm'}\Gamma(\Omega_{m'})J_{m}(2\eta)J_{m'}(2\eta)
\bigl\{\bigl(1+\bar n(\Omega_{m'})\bigr)[R_{eg},R_{ge}\rho]e^{i(m-m')(\omega t + \phi)} \nonumber \\
&+& \bar n(\Omega_{m'})[R_{ge},R_{eg}\rho]e^{-i(m-m')(\omega t + \phi)}\bigr \} +H.c., \label{rpn}
\end{eqnarray}
where we made the notations: $\Omega_{m}=\Omega + m\omega$, with $|m\omega| \ll \Omega$, $2\Gamma(\Omega_{m})
=\pi\sum_{p}g^{2}_{p}\delta(\omega_{p}-\Omega_{m})$, and $\bar n(x) =1/[\exp{(\hbar x/k_{B}T)}-1]$ being the mean phonon 
number at the frequency $x$ and temperature $T$, while $k_{B}$ is the Boltzmann constant.

Respectively, the resonator's mode interaction with its surrounding thermostat is described as usual, i.e., 
\begin{eqnarray}
\dot \rho_{pt} &=& -\frac{\kappa}{2}\bigl(1 + \bar n_{r}\bigr)[a^{\dagger},a\rho] 
-\frac{\kappa}{2}\bar n_{r}[a,a^{\dagger}\rho] + H.c.. \label{rpt}
\end{eqnarray}
Here $\bar n_{r} =1/[\exp{(\hbar \omega_{r}/k_{B}T)}-1]$ is the corresponding mean photon number filling the resonator 
mode due to its interaction with the environmental thermostat, while $\kappa$ is the cavity photon decay rate.  The 
dephasing of the cavity photon mode, with a rate $\tilde \kappa$, is incorporated via the following contribution to 
the final master equation, see e.g. \cite{wm},
\begin{eqnarray}
\dot \rho_{dp} = -\frac{\tilde\kappa}{2}\bigl[a^{\dagger}a,a^{\dagger}a\rho\bigr] + H.c..
\label{dpg}
\end{eqnarray}

Thus, the master equation describing the entire system is represented finally as
\begin{eqnarray}
\dot \rho + \frac{i}{\hbar}\bigl[\bar H,\rho\bigr] = \Lambda \rho, \label{eqf}
\end{eqnarray}
where $\Lambda \rho$ is given by the sum of all the right-side parts of Eqs.~(\ref{reL} - \ref{dpg}), respectively. The 
Hamiltonian entering in the Eq.~(\ref{eqf}) is given then as follows
\begin{eqnarray}
\bar H = -\hbar\bar g_{r}\bigl(a^{\dagger}R_{ge} + R_{eg}a\bigr), \label{hib}
\end{eqnarray}
where we have considered that resonances occur at $\omega_{r}=\Omega \pm |m\omega|$, $|m| \ge 1$ so that 
$|m\omega| \ll \Omega$, whereas $\bar g_{r}=g_{r}\sin{\theta}J_{m}(2\eta)$. All other time-dependent terms were 
omitted as being rapidly oscillating. Notice that the cross-correlations between the DQD, photon and phonon 
decay channels, were omitted here as being of higher order \cite{tam1}.

In the next section, based on the master equation (\ref{eqf}), we obtain the equations of motion describing the 
combined DQD and the resonator mode degrees of freedom, respectively, and present the steady-state expectation 
values of quantities of interest.

\section{The equations of motion \label{seqm}}
In the following, using the master equation (\ref{eqf}) derived in the previous section, we write down the corresponding 
system of equations of motion, obtained within the performed approximations, describing the entire sample incorporating 
a DQD coupled respectively to the leaking cavity boson-mode, i.e.,
\begin{eqnarray}
\dot P^{(0)}_{n}&=& - (\Gamma_{Ls}+\Gamma_{Lc})P^{(0)}_{n} + \Gamma_{Rc}P^{(1)}_{n}+\Gamma_{Rs}P^{(2)}_{n} - 
\kappa(1+\bar n_{r})\bigl(nP^{(0)}_{n} - (n+1)P^{(0)}_{n+1}\bigr) \nonumber \\
&-& \kappa\bar n_{r}\bigl((n+1)P^{(0)}_{n} - nP^{(0)}_{n-1}\bigr), \nonumber \\
\dot P^{(1)}_{n}&=&\Gamma_{Ls}P^{(0)}_{n} - (\Gamma^{(-)} + \Gamma_{Rc})P^{(1)}_{n} + \Gamma^{(+)}P^{(2)}_{n} 
+ i\bar g_{r}P^{(3)}_{n} - \kappa(1+\bar n_{r})\bigl(nP^{(1)}_{n} - (n+1)P^{(1)}_{n+1}\bigr) \nonumber \\
&-& \kappa\bar n_{r}\bigl((n+1)P^{(1)}_{n} - nP^{(1)}_{n-1}\bigr), \nonumber \\
\dot P^{(2)}_{n}&=& \Gamma_{Lc}P^{(0)}_{n} + \Gamma^{(-)}P^{(1)}_{n} - (\Gamma^{(+)} + \Gamma_{Rs})P^{(2)}_{n} 
+ i\bar g_{r}P^{(5)}_{n} - \kappa(1+\bar n_{r})\bigl(nP^{(2)}_{n} - (n+1)P^{(2)}_{n+1}\bigr) \nonumber \\
&-& \kappa\bar n_{r}\bigl((n+1)P^{(2)}_{n} - nP^{(2)}_{n-1}\bigr), \nonumber \\
\dot P^{(3)}_{n}&=& 2i\bar g_{r}n(P^{(1)}_{n} - P^{(2)}_{n-1}) - \Gamma_{r}P^{(3)}_{n}/2 -\kappa(1+\bar n_{r}) 
\bigl((2n-1)P^{(3)}_{n} - 2(n+1)P^{(3)}_{n+1}-2P^{(5)}_{n}\bigr)/2 \nonumber \\
&-& \kappa\bar n_{r}\bigl((2n+1)P^{(3)}_{n} - 2nP^{(3)}_{n-1}\bigr)/2, \nonumber \\
\dot P^{(4)}_{n}&=& - \Gamma_{r}P^{(4)}_{n}/2 -\kappa(1+\bar n_{r})\bigl((2n-1)P^{(4)}_{n} - 2(n+1)P^{(4)}_{n+1} 
+ 2P^{(6)}_{n}\bigr)/2 \nonumber \\
&-& \kappa\bar n_{r}\bigl((2n+1)P^{(4)}_{n} - 2nP^{(4)}_{n-1}\bigr)/2, \nonumber \\
\dot P^{(5)}_{n}&=& 2i\bar g_{r}(n+1)(P^{(2)}_{n} - P^{(1)}_{n+1}) - \Gamma_{r}P^{(5)}_{n}/2 - \kappa(1+\bar n_{r})
\bigl((2n+1)P^{(5)}_{n} - 2(n+1)P^{(5)}_{n+1}\bigr)/2 \nonumber \\
&-& \kappa\bar n_{r}\bigl((2n+3)P^{(5)}_{n} - 2nP^{(5)}_{n-1} + 2P^{(3)}_{n}\bigr)/2, \nonumber \\
\dot P^{(6)}_{n}&=&-\Gamma_{r}P^{(6)}_{n}/2 - \kappa(1+\bar n_{r})\bigl((2n+1)P^{(6)}_{n}- 2(n+1)P^{(6)}_{n+1}\bigr)/2 
- \kappa\bar n_{r}\bigl((2n+3)P^{(6)}_{n} \nonumber \\
&-& 2nP^{(6)}_{n-1} - 2P^{(4)}_{n}\bigr)/2. \label{eqm}
\end{eqnarray}
The system of equations (\ref{eqm}) can be easily obtained using the master equation (\ref{eqf}), if one first gets the 
equations of motion for $\rho_{\alpha\beta}$=$\langle \alpha|\rho|\beta \rangle$, $\{ \alpha,\beta \in e, g, 0\}$, see 
also \cite{tqf}, and then writing the corresponding equations for the following variables: $\rho^{(0)}=\rho_{00}$, 
$\rho^{(1)}=\rho_{gg}$, $\rho^{(2)}=\rho_{ee}$, $\rho^{(3)}=a^{\dagger}\rho_{eg} - \rho_{ge}a$, 
$\rho^{(4)} = a^{\dagger}\rho_{eg} + \rho_{ge}a$, $\rho^{(5)}=a\rho_{ge} - \rho_{eg}a^{\dagger}$ and 
$\rho^{(6)}=a\rho_{ge} + \rho_{eg}a^{\dagger}$. The projection on the Fock states $|n\rangle$, i.e. 
$P^{(j)}_{n}=\langle n|\rho^{(j)}|n\rangle$, with $j \in \{0, \cdots, 6\}$ and $n \in \{0,\infty\}$, will lead
us to the system of equations (\ref{eqm}). There, 
\begin{eqnarray*}
\Gamma^{(+)}&=& \sin^{2}{\theta}\sum_{m}\bigl(1+\bar n(\Omega_{m})\bigr)\Gamma(\Omega_{m})J^{2}_{m}(2\eta), 
\nonumber \\
\Gamma^{(-)} &=& \sin^{2}{\theta}\sum_{m}\bar n(\Omega_{m})\Gamma(\Omega_{m})J^{2}_{m}(2\eta), 
\end{eqnarray*}
while
\begin{eqnarray*}
\Gamma_{ls}&=& \sin^{2}{(\theta/2)}\sum_{m}\Gamma_{l}(\Omega'_{m})J^{2}_{m}(\eta), \nonumber \\
\Gamma_{lc}&=&\cos^{2}{(\theta/2)}\sum_{m}\Gamma_{l}(\Omega'_{m})J^{2}_{m}(\eta), \nonumber \\
\Gamma_{r} &=& \Gamma^{(+)} + \Gamma^{(-)} + \Gamma_{Rs} + \Gamma_{Rc} + \tilde\kappa,
\end{eqnarray*}
with $\{l \in R, L\}$ and we have assumed also that $\omega \gg \{\Gamma^{(\pm)},\Gamma_{ls},\Gamma_{lc}\}$, respectively.

Actually, to solve the infinite system of equations (\ref{eqm}), one truncates it at a certain maximum value $n=n_{max}$ so 
that a further increase in its value, i.e. $n_{max}$, does not modify the obtained results. As a consequence, the steady-state 
cavity-mode photon mean number, i.e. $\langle m\rangle =\langle a^{\dagger}a\rangle$, and its squared expectation 
value, that is $\langle m^{2}\rangle =\langle (a^{\dagger}a)^{2}\rangle$, are expressed as:
\begin{eqnarray}
\langle m\rangle = \sum^{n_{max}}_{n=0}n\bigl(P^{(0)}_{n} + P^{(1)}_{n} + P^{(2)}_{n}\bigr), \label{nmf}
\end{eqnarray}
and
\begin{eqnarray}
\langle m^{2}\rangle = \sum^{n_{max}}_{n=0}n^{2}\bigl(P^{(0)}_{n} + P^{(1)}_{n} + P^{(2)}_{n}\bigr), \label{nmfp}
\end{eqnarray}
with
\begin{eqnarray*}
\sum^{n_{max}}_{n=0}\bigl(P^{(0)}_{n}+ P^{(1)}_{n}+P^{(2)}_{n}\bigr)=1.
\end{eqnarray*}
\begin{figure}
\centering
\includegraphics[width =8cm]{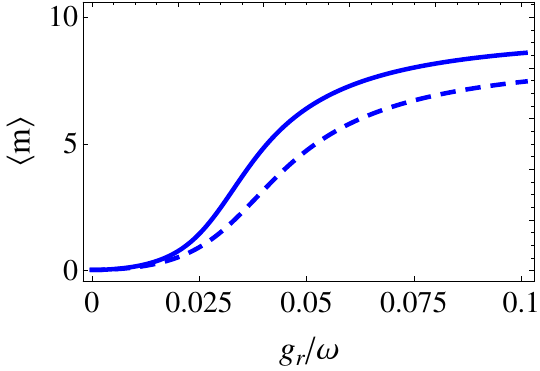}
\caption{\label{fig-1}
The steady-state cavity mean photon number $\langle m\rangle=\langle a^{\dagger}a\rangle$ as a function of $g_{r}/\omega$. 
Here, $\tau/\epsilon=0.2$, $\omega/\epsilon=0.1$, $\epsilon_{0}/(2\omega)=0.3$, $\Gamma_{L}/\omega=\Gamma_{R}/\omega=0.03$, 
$\kappa/\omega=\tilde\kappa/\omega=10^{-3}$, $\hbar\epsilon/(k_{B}T)=5$, and $\bar n_{r}=0.019$. The solid line is plotted for 
$\Gamma/\omega=0$, while the dashed one for $\Gamma/\omega=0.02$, respectively.}
\end{figure}

Respectively, the Fano's factor is given by the following expression \cite{wm},
\begin{eqnarray}
F = \bigl(\langle m^{2}\rangle - \langle m\rangle^{2}\bigr)/\langle m\rangle. \label{ff}
\end{eqnarray}
Values of the Fano's factor below unity, i.e. $F < 1$, mean that squeezed photon number states are generated possessing 
sub-Poissonian photon statistics. Additional characteristics of the cavity photons statistics can be inferred from the steady-state 
normalised second- and third-order photon correlation functions, defined in the usual way \cite{gb1,gb2}, namely,
\begin{eqnarray}
g^{(2)}(0) &=& \frac{\langle a^{\dagger 2}a^{2}\rangle}{\langle m\rangle^{2}} \nonumber \\
&=&\sum^{n_{max}}_{n=0}n(n-1)\bigl(P^{(0)}_{n}+ P^{(1)}_{n}+P^{(2)}_{n}\bigr)/\langle m\rangle^{2}, \label{crf2}
\end{eqnarray}
while
\begin{eqnarray}
g^{(3)}(0) &=& \frac{\langle a^{\dagger 3}a^{3}\rangle}{\langle m\rangle^{3}} \nonumber \\
&=& \sum^{n_{max}}_{n=0}n(n-1)(n-2)\bigl(P^{(0)}_{n}+ P^{(1)}_{n}+P^{(2)}_{n}\bigr)/\langle m\rangle^{3}, \label{crf3} 
\end{eqnarray}
respectively. Notice that, $g^{(2)}(0)<1$ characterizes sub-Poissonian, $g^{(2)}(0)>1$ super-Poissonian, and $g^{(2)}(0)=1$ 
Poissonian photon statistics. Furthermore, the condition $g^{(3)}(0) < g^{(2)}(0) <1$ means that the generated cavity EMF 
consists of a beam of non-classical single photons.

In the following section, we shall describe the resonator's photon quantum dynamics, via Exps.~(\ref{nmf}-\ref{crf3}), 
based on the numerical solution of the system of equations (\ref{eqm}).
\begin{figure}
\centering
\includegraphics[width =8cm]{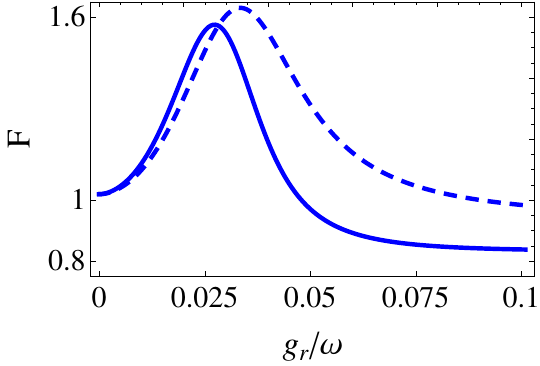}
\caption{\label{fig-2}
The steady-state behaviour of the Fano's factor $F$ as a function of $g_{r}/\omega$. 
The involved parameters are as in Fig.~(\ref{fig-1}).}
\end{figure}

\section{Results and Discussion \label{RD}}
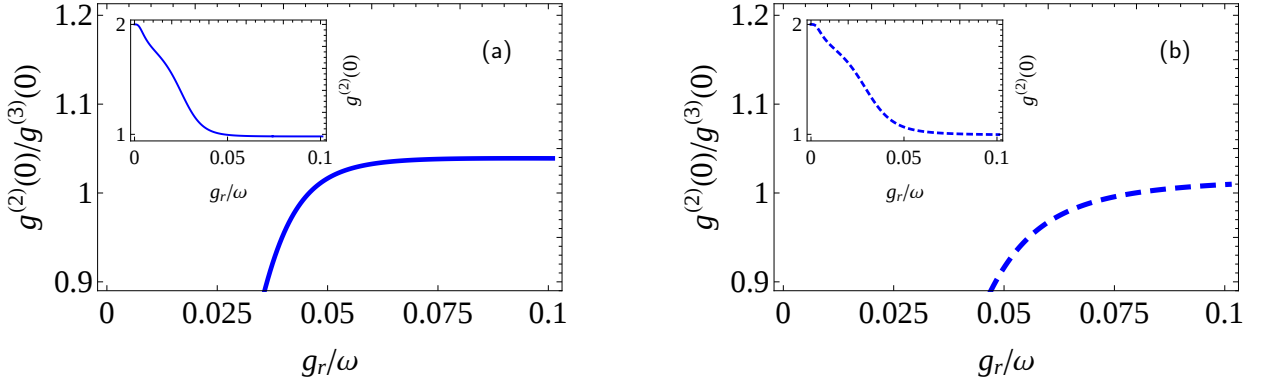
\begin{figure}
\centering
\begin{subfigure}{0.45\textwidth}
\includegraphics[width=\linewidth]{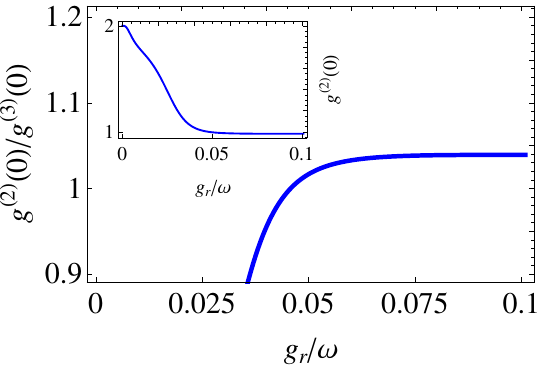}
\end{subfigure}
\hfill
\begin{subfigure}{0.45\textwidth}
\includegraphics[width=\linewidth]{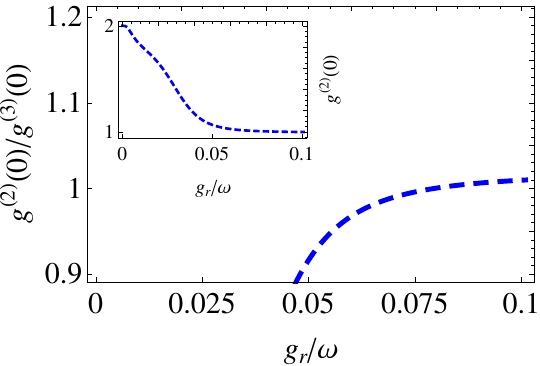}
\end{subfigure}
\begin{picture}(0,0)
\put(-290,120){(a)}
\put(-35,120){(b)}
\end{picture}
\caption{\label{fig-3} The ratio of the steady-state values of correlation functions $g^{(2)}(0)/g^{(3)}(0)$ as a function 
of $g_{r}/\omega$, for (a) $\Gamma/\omega=0$ while (b) $\Gamma/\omega=0.02$, respectively. The inset shows 
the corresponding steady-state behaviour of the second-order photon-photon correlation function. The other involved 
parameters are as in Fig.~(\ref{fig-1}).}
\end{figure}

We proceed by showing the cavity mean photon number in the steady-state, depicted in Fig.~(\ref{fig-1}), for some 
generic parameters of interest and within the performed approximations. We suppose that the cavity resonance 
occurs at $\omega_{r}=\Omega \pm \omega$, see Fig.~(\ref{fig-0}), and therefore $\bar g_{r}=\pm g_{r}\sin{\theta}J_{1}(2\eta)$ 
for $m=\pm 1$, respectively, although both situations lead to the same results. It is assumed that the electronic 
reservoirs are flat, at the qubit's frequency, so that the corresponding electronic pumping rates are 
$\Gamma_{L}(\Omega'_{m})\equiv\Gamma_{L}$ and $\Gamma_{R}(\Omega'_{m}) \equiv \Gamma_{R}$, respectively. 
Since $\sum^{\infty}_{m=-\infty}J^{2}_{m}(\eta)=1$, one has that $\Gamma_{ls}=\sin^{2}(\theta/2)\Gamma_{l}$ and 
$\Gamma_{lc}=\cos^{2}(\theta/2)\Gamma_{l}$, with $l \in \{L,R\}$. The phonon reservoir is considered flat too at 
the qubit's frequency and, thus, $\Gamma(\Omega_{m}) \equiv \Gamma$. The entire sample is kept at lower 
environmental temperatures. As a result, the mean cavity photon number starts increasing from $\bar n_{r}$ as a 
function of the ratio $g_{r}/\omega$ and then reach the saturation, see Fig.~(\ref{fig-1}). Interestingly here, the 
Fano's factor has values lower than unity, meaning that squeezed cavity photon number states are generated 
obeying the sub-Poissonian photon statistics, see Fig.~(\ref{fig-2}) and the insets in Fig.~(\ref{fig-3}). Furthermore, 
to further characterise the photon features, in Fig.~(\ref{fig-3}) we plot the ratio $g^{(2)}(0)/g^{(3)}(0)$ against 
$g_{r}/\omega$. One can observe that this ratio is larger than unity, while $g^{(2)}(0)<1$, for some experimentally 
accessible ranges of $g_{r}/\omega$, denoting that the generated cavity EMF consists of a stream of single microwave 
photons possessing quantum features. However, the described situation is quite sensitive on the presence of phonons 
in the sample, compare the solid and dashed curves in Fig.~(\ref{fig-2}) and Fig.~(\ref{fig-3}). Actually, in the 
good cavity limit, the quantum photon features persist for lower temperatures and lower phonon rates compared 
to electronic pumping ones, that is, $\Gamma \ll \{\Gamma_{L},\Gamma_{R}\} \ll \omega$. The cavity mode dephasing 
$\tilde\kappa$ has to be less or of the same order as cavity damping rate $\kappa$. Notice that if $g_{r}=0$, then 
$g^{(2)}(0)=2$ while $g^{(3)}(0)=6$, i.e., one has a thermal distribution of cavity photons in this particular case, 
namely, for $T\not=0$, where Fano's factor equals to $F=1 + \bar n_{r}$. 
Finally, we considered experimentally achievable parameters \cite{ppr1,ppr2}, describing a leaking resonator 
having the frequency $\omega_{r} \sim 1{\rm GHz}$, decay rate $\kappa \sim 1{\rm MHz}$ and DQD-cavity coupling 
strengths up to $g_{r} \sim 1{\rm GHz}$, embedded in a ${\rm mK}$ temperature environment. Furthermore, the involved rates 
were selected within the secular approximation ($\Gamma_{L/R}$ being of the order of several ${\rm MHz}$, see e.g. 
\cite{ppr1}), that is, smaller than the qubit frequency $\Omega$, in accordance with the validity of the master equation (\ref{eqf}).
The results are robust against adding an additional state, where both dots are being occupied by an electron, within the performed 
approximations, see e.g. \cite{tam1,tam2}.

The advantage of the proposed model for generation of a photon flux, formed from single microwave photons 
possessing quantum features, consists in its flexibility. This means that the photons stream can be generated at 
various frequencies, depending on a concrete resonator, by simply modulating the DQD's frequency to match 
the cavity one. Additionally, the photon quantum properties occur at weaker DQD-resonator coupling strengths
compared to a similar setup, but in the absence of frequency modulation.  Furthermore, the phonon influence can 
be weakened if one engineers the phonon reservoir, namely, the phonon bandwidth is not being considered  flat 
but bounded by some frequency range. Particularly, the phonon rates $\Gamma^{(\pm)}$ depend on the external 
control parameter $\eta$ through $J^{2}_{m}(2\eta)$. The squared Bessel function, i.e. $J^{2}_{m}(2\eta)$ with 
$m=0$, for instance, meaning that the phonon modes ale localized around qubit's frequency $\Omega$, can be 
selected small enough, via an appropriate choice of $\eta$, in order to make the phononic rates $\Gamma^{(\pm)}$ 
negligible. 

\section{Summary \label{sum}}
Summarising, we have investigated the quantum dynamics of a double quantum dot two-level emitter coupled to the leaking 
mode of a microwave resonator. Under strong Coulomb interaction limit and Born-Markov approximations with respect to
the involved electronic, phononic or photonic reservoirs, we derived the corresponding master equation where the qubit's 
frequency is being additionally time-modulated. The frequency of the external applied field, which modulates the qubit 
energy separation, is considered bigger than the electronic pumping rates or phonon decay rates, respectively. As a 
consequence, we have demonstrated that the output cavity electromagnetic field consists of a stream of single microwave 
photons obeying the sub-Poissonian statistics. The reported result is sensitive on the environmental temperatures which 
have to be kept low enough. The presence of phonons in the system affect the photon quantum nature too, however, their 
influence can be weakened via phonon reservoir engineering. The benefits from the proposed model rely in the fact that 
the photon flux, possessing quantum features, can be generated at resonator's microwave frequencies symmetrically 
located around the qubit one, i.e. $\omega_{r}=\Omega \pm m\omega$, $\{m \in 1,2 \cdots\}$.


\section{Acknowledgments}
We highly appreciate the financial support from the Romanian Ministry of Education and Research via the 
PN-IV-P8-8.3-ROMD-2023-0013 research grant, within PNCDI IV. MM is grateful for the nice hospitality 
of the Department of Theoretical Physics at the Horia Hulubei National Institute of Physics and Nuclear 
Engineering as well as for partial support from the Moldavian Ministry of Education and Research, through 
grant No. 011205.

\end{document}